\documentstyle[12pt,twoside]{article}
\normalsize
\begin{document}
\bibliographystyle{plain}
\begin{titlepage}
\begin{flushright}
UWThPh-1996-24 \\
March 1996
\end{flushright}
\vspace{2cm}
\begin{center}
{\Large \bf Need for Two Vectorlike Families in SUSY Composite Models}
{\large ${}^\dagger$} \\[40pt]
{\bf H. Stremnitzer} \\
{\small {\it e-mail: strem@pap.univie.ac.at}} \\[10pt]
Institute for Theoretical Physics \\
University of Vienna, Vienna, Austria\\
\vfill
{\bf Abstract} \\
\end{center}
  
Within the context of a viable and economical SUSY preon model,
two vector--like families $Q_{L,R} = (U,D,N,E)_{L,R}$ and
$Q^\prime_{L,R} = (U^\prime,D^\prime,N^\prime,E^\prime)_{L,R}$ with 
masses of order 1 TeV, one of which is a doublet of $SU(2)_L$ and the 
other a doublet of $SU(2)_R$, have been predicted to exist together with 
the three observed chiral families.
The existence of these two vector--like families turns out to be 
crucial especially for an explanation of the inter--family 
mass--hierarchy and therefore for the SUSY preon model itself.
This paper is devoted to a detailed study of the expected masses, 
mixings and decay modes of the vector--like fermions.
Including QCD renormalization--effects, the masses 
of the vector--like quarks are expected to lie in the range of 500 GeV to 
about 2.5 TeV, while those of the vector--like leptons are expected to 
be in the range of 200 GeV to 1 TeV.

\vfill \noindent
{\small
${}^\dagger$ {\it Invited talk, given at the Hadron Structure '96
Meeting in Stara Lesna, Slovakia, Feb. 12-16, 1996}}

\end{titlepage}

{\bf 1. Generalities} \\

    Recently, it has been shown [1-3] that the idea that quarks,
leptons, and Higgs bosons are composites and that their
constituents possess local supersymmetry can be realized in the
context of a viable and economical preon model which has
many attractive features. These include:
\begin{itemize}
\item Understanding of the quark and lepton mass matrix,
including KM-Matrix and $CP$-violation [2];
\item Understanding family replication [3];
\item Cosmological features [4].
\end{itemize}

    In this talk, I would like to concentrate on a specific
feature of this preon model, namely the appearance of two
families of vector--like quarks and leptons  $Q_{L,R}$ and
$Q^\prime_{L,R}$ in the TeV range [5].
Let me first recall the {\it essential role} which 
these two families play in providing an explanation of the inter--family 
mass hierarchy. Since $Q_L$ and $Q_R$ couple symmetrically 
to $SU(2)_L \times U(1)_Y$ gauge bosons, the mass term 
$(\overline{Q}_LQ_R+h.c)$ and likewise $(\overline{Q}_L^{\prime}
Q_R^{\prime}+h.c)$ preserve $SU(2)_L \times U(1)_Y$.
This, however, infers that the oblique parameters $S,T$ and $U$ [6],
do not receive contributions from these vector--like families, in the leading 
approximation.  As a result, the prevailing set of measurements of the 
electroweak parameters, despite their precision, are not sensitive enough 
to the existence of vector--like families [7] -  unlike the case of a fourth 
chiral family which 
is slightly disfavored by the measurement of the $S,T$ and 
$U$--parameters. One therefore tends to believe that
if new families beyond the three chiral ones are yet to be 
found, they are more likely to exhibit vectorial rather than chiral 
couplings to $W_L$'s and $W_R$'s.

     The model under consideration is based on a set of
masseless chiral superfields, each belonging to the
fundamental representation of the metacolor gauge symmetry
$SU(N)$. The superfields carry also flavor-color quantum numbers
according to the gauge group $SU(4)_c \times SU(2)_L\times SU(2)_R$.
It is assumed that the metacolor force becomes strong and
confining at a scale $\Lambda_M \simeq 10^{11}$ GeV, with the
following effects:
\begin{itemize}
\item Three light chiral families of composite quarks and
leptons $\left( q_{L,R}^{i}\right)_{i=1,2,3}$  and two 
 vector-like families $Q_{L,R}$ and $Q^{\prime}_{L,R}$, coupling
 vectorially to $W_{L}$'s and $W_{R}$'s, are formed;
\item Supersymmetry-breaking condensates are formed; they include
 the metagaugino condensate
$\langle\vec{\lambda}\cdot\vec{\lambda}\rangle$ and the matter
fermion-condensates $\langle\bar{\psi}^{a}\psi^{a}\rangle\,$. 
  Noting that, within the class of models under consideration, the 
index theorem prohibits a dynamical breaking of supersymmetry
in the absence of gravity [8], so
the formation of these condensates must need the collaboration between
the metacolor force and gravity.  As a result, each of these condensates is
expected to be damped by one power of $\left( \Lambda_{M} /
M_{P\ell}\right) \simeq 10^{-8}$ relative to $\Lambda_{M}\,$ [9]:
\begin{eqnarray}
\langle\vec{\lambda}\cdot\vec{\lambda}\rangle & = &
\kappa_{\lambda}\Lambda_{M}^{3}\left(\Lambda_{M} /
M_{P\ell}\right) , \nonumber \\*
\langle\bar{\psi}^{a}\psi^{a}\rangle & = &
\kappa_{\psi_{a}}\Lambda_{M}^{3}\left(\Lambda_{M} / M_{P\ell}\right)
\end{eqnarray}
Here, the indices $a$ are running over color and flavor quantum numbers.
 The condensates $\langle\bar{\psi}^{a}\psi^{a}\rangle\,$,
 break not only SUSY but also the electroweak
symmetry $SU(2)_{L}$ $\times$ $U(1)_{Y}\,$ therefore giving mass
to the electroweak gauge bosons. The coefficients $\kappa_{\lambda}$ 
and $\kappa_{\psi_{a}}\,$, apriori, are expected
to be of order unity within a factor of ten (say), although
$\kappa_{\lambda}$ is expected to be bigger than $\kappa_{\psi}$'s, 
typically by factors of 3 to 10, because the $\psi$'s are in the fundamental
and the $\lambda$'s are in the adjoint representation of the metacolor
group.
\item Furthermore, supersymmetry-preserving condensates, which
however break the gauge group $SU(4)_c \times SU(2)_L\times SU(2)_R$
to the low-energy gauge group $SU(3)_c \times SU(2)_L \times U(1)_Y$
are assumed to form as well. They provide a large superheavy
Majorana mass to the right-handed neutrinos and may play an
interesting role in the discussion of inflationary models [4].
\end{itemize}

Now, the vector-families $Q_{L,R}$ and
$Q_{L,R}^{\prime}$ acquire relatively heavy masses through the
metagaugino condensate $\langle\vec{\lambda}\cdot\vec{\lambda}\rangle$
of order $\kappa_{\lambda}\Lambda_{M}\left(\Lambda_{M} / M_{P\ell}\right)$
$\sim$ 1~TeV which are independent of flavor and color.  But
the chiral families $q_{L,R}^{i}$ acquire masses primarily through their
mixings with the vector-like families $Q_{L,R}$ and $Q_{L,R}^{\prime}$
which are induced by $\langle\bar{\psi}^{a}\psi^{a}\rangle\,$.  This is
because the direct mass-terms  cannot be induced through
two-body condensates. Thus, ignoring $QCD$
corrections and higher order condensates for a moment, 
the Dirac-mass matrices of all four types -
i.e., up, down, charged lepton, and neutrino - have the form:
\begin{eqnarray}
M^{(o)} =
\bordermatrix{& q_{L}^{i} & Q_{L} & Q_{L}^{\prime}\cr
\overline{q_{R}^{i}} & O & X\kappa_{f} & Y\kappa_{c}\cr
\overline{Q_{R}} & Y^{\prime\dagger}\kappa_{c} & \kappa_{\lambda} & O\cr
\overline{Q_{R}^{\prime}} & X^{\prime\dagger}\kappa_{f} & O &
\kappa_{\lambda}\cr}~~~.
\end{eqnarray}
Here, the index $i$ runs over three families, $f,c$ denotes
flavor- or color-type condensate, and
the quantitites  $X\,$, $Y\,$, $X^{\prime}$ and
$Y^{\prime}$ are column matrices in the family-space and have
their origin in the detailled vertex structure of the
corresponding preonic diagrams. They are expected to be numbers 
of order $\simeq 1 $. As a
result, the Dirac mass-matrices of all four types have a natural
see-saw structure. \\
 
{\bf 2. Discussion of the Mass Matrix} \\

A detailled discussion of the mass matrix can be summarized as follows:
\begin{itemize}
\item In the absence of electroweak corrections,
left-right symmetry and flavor-color independence of the metacolor force
guarantee (a) $X = X'$ and $Y = Y'$, and (b)
the {\it same} $X,~Y$ and $\kappa_\lambda$ enters into up, down, lepton,
or neutrino-type matrix.
This results in an enormous reduction of parameters. 
The whole $5 \times 5$ mass-matrices of the four types is
essentially determined by just six effective parameters.
\item Since the rank of the matrix is four, one family (the electron-family),
is strictly massless to all orders. The essential parameters, however,
can be adjusted to give reasonable masses for the second and third
chiral family. Inclusion of electroweak corrections (five more 
parameters of order 10\% ) do not change the rank, however, they 
give a possibility to fit the KM-Matrix elements.
\item Also, $CP$-violating phases can be rotated away in the
above matrix. The inclusion of electroweak corrections does not
alter this. In order to gain mass for the electron family, a small
contribution from 4-body condensates has to be included into the
mass entries for the chiral quarks. These contributions are
damped by $(\Lambda_{M} / M_{P\ell})^2$ and yield masses of
order $\simeq 1MeV$. Interesting enough, they also provide a nonzero
phase in the KM Matrix. Therefore the interesting conclusion in
this model is, that the mass of the first chiral family and
$CP$-Violation has the same origin.
\end{itemize}

We therefore can summarize that the above fermion mass matrix including
radiative corrections on the metacolor scale as well as
renormalization group corrections and the appearance of small
direct mass entries can give a reasonable understanding of all
fermion masses, mixing angles, and $CP$-phases. Naturally, the
key ingredient in this prediction is the seesaw mechanism,
provided by the existence of the two set of vector--like quarks.
The prediction of the masses, coupling constants and decay rates
of these quarks and their verification by experiments is 
therefore a crucial test of the model itself. \\

{\bf 3. Properties of the heavy vector--like quarks} \\

Heaving fixed our parameters by the known chiral families of
quarks and leptons, we now start to make detailled predictions
about the behavior of the $Q$-family (the $SU(2)_L$-doublet) and
the $Q^{\prime}$-family (which is an $SU(2)_L$ singlet). Obviously,
the off-diagonal entries in the mass matrix yield also a mixing
between those two families in second order seesaw, yielding 
four mass eigenstates $(U_1, D_1, E_1, N_1)_{L,R}$ (mostly weak
doublets), and ($U_2, D_2, E_2, N_2)_{L,R}$ (mostly singlets).
Furthermore, we have to calculate renormalization group corrections
for the running masses from $\Lambda_M$, where the above matrix
was defined, down to about 1.5TeV, which is the expected
range of the masses of
the heavy quarks. In case of $QCD$, this correction is as large
as a factor 2.39, whereas the electroweak correction factors
vary between $1.001 - 1.23$ for the different mass values.
Putting things together, we arrive at the following list of
mass values for a representative set of parameters:
\begin{eqnarray}
\kappa_\lambda & \approx & 575~GeV,~\kappa_u/\kappa_\lambda = 1/6,~
\kappa_d/\kappa_u \simeq 1/40 \nonumber \\
& ~ & \kappa_r/\kappa_\lambda \approx 1/3,~
\kappa_l/\kappa_\lambda \approx 1/3 
\end{eqnarray}
This yields 
\begin{eqnarray}
(m_t, m_b, m_\tau)_{1.5~TeV} \simeq (135, 3.4, 1.5)~GeV \nonumber \\
(m_t, m_b, m_\tau)_{phys} \simeq (157, 4.7, 1.7)~GeV 
\end{eqnarray}
\begin{eqnarray}
U_1 \simeq 1814~ GeV ~~~~~N_1 \simeq 765~GeV \nonumber \\
D_1 \simeq 1787 ~GeV ~~~~~E_1 \simeq 763~GeV \nonumber \\
U_2 \simeq 1504 ~GeV ~~~~~N_2 \simeq 581~GeV \nonumber \\
D_2 \simeq 1465 ~GeV ~~~~~E_2 \simeq 667~GeV 
\end{eqnarray}

While the precise values of the masses and the mixing angles depend upon 
the specific choice of the parameters, a few qualitative features 
would still remain which are worth noting:
\begin{itemize}
\item The masses of the heavy neutrinos are Dirac masses. We rely
here in a scenario [10] where the heavy neutrinos do not acquire a superheavy
Majorana mass, in contrast to the chiral righthanded neutrinos $\nu^i_R$.
\item The smallness of the mixing angles from second order seesaw implies
that $D_1$ and even $U_1$ are mostly composed of $Q$--fermions which are
$SU(2)_L$--doublets while $U_2$ and $D_2$ are mostly composed of 
$Q^\prime$--fermions which are $SU(2)_R$--doublets.  This is important 
for their decay modes.  
\item Note that the pair $U_1$ and $D_1$ are nearly degenerate to within 
about 10-30 GeV, so also the pair $N_1$ and $E_1$, and to a lesser
extent the pair
$N_2$ and $E_2$.  But the $(U_1,D_1)$ pair is substantially heavier, by 
about a few hundred GeV, than the pair $(U_2,D_2)$.  Similarly the
$(N_1,E_1)$ pair is heavier by about 100 GeV than the pair $(N_2,E_2)$.  
This is because the $(U_1,D_1)$ and also the $(N_1,E_1)$ pair receive 
enhancement due to a $SU(2)_L$--renormalization factor, which is, 
however, absent for the $(U_2,D_2)$ and $(N_2,E_2)$--pairs.  
\item Given this mass--pattern, we see that $U_1 \rightarrow D_1 + W$ and 
likewise $N_1 \rightarrow E_1 + W$ are forbidden kinematically, while 
decays such as $U_1 \rightarrow D_2 +W,~U_1 \rightarrow U_2+Z,~
D_1 \rightarrow U_2+W,~D_1 \rightarrow D_2+Z$ and possibly
$U_2 \rightarrow D_2+W$ are kinematically allowed.  
\end{itemize}

The most dominant decay modes (with rates $\Gamma$ up to 100 GeV)
are the decays $U_1 \rightarrow U_2 + Z$, $U_2 \rightarrow b + W^+$,
$D_1 \rightarrow U_2 + W^-, t + W^-$, and $D_2 \rightarrow t + W^-$.
For the leptons, the decays of $E_1, E_2 $ into $\tau + Z $ are 
dominant, but highly suppressed, indicating a small decay rate,
whereas the decays of the neutrinos into chiral leptons and
$Z$ or $W$ is appreciable.

Production of the heavy quarks in pairs by hadronic colliders at SSC and LHC 
energies has been studied in a number of papers [11].  These studies 
typically yield production cross sections of $3.10^{-4}$ nb at $\sqrt{s}=
40~TeV$. Assuming that a future version of the SSC will be built one day 
in the near future, the production cross section noted above would lead 
to about $2.5 \times 10^4$ events per year for $m_U=1~TeV$, with a
luminosity of $10^{33} cm^{-2}s^{-1}$.  
 
In summary, two vector--like families, not more not less, with one 
coupling vectorially to $W_L$'s and the other to $W_R$'s (before 
mass--mixing), with masses of order $\simeq 1TeV$, constitute a 
{\it hall--mark} and a crucial prediction of our SUSY preon
model [1-3]. There does not seem to be any other model including 
superstring--inspired models of elementary quarks and leptons 
which have a good reason to predict two such complete 
vector--like families with masses in the TeV range.\\

{\bf References} \\
\newcounter{000}~~
\begin{list}{[\arabic{000}]}{\usecounter{000}
\labelwidth=1cm\labelsep=.5cm}
\item J. C. Pati, {\it Phys. Lett.} {\bf B228} (1989) 228;
\item{K. S. Babu, J. C. Pati, H. Stremnitzer,} {\it Phys. Rev.
Lett.} {\bf 67} ({1991}) {1688};
\item{K. S. Babu, J. C. Pati, H. Stremnitzer}, {\it Phys. Lett.} 
{\bf B256} (1991) {206};
\item{M. Cveti\v{c}, T. H\"ubsch, J. C. Pati, H. Stremnitzer},
{\it Phys. Rev.} {\bf D40} (1989) {1311};
\item{K. S. Babu, J. C. Pati, H. Stremnitzer}, {\it Phys. Rev.} 
{\bf D51} (1995) {2451};
\item{M. Peskin, T. Takeuchi}, {\it Phys. Rev. Lett.}
{\bf 65} (1990) {964};
\item{K. S. Babu, J.C. Pati, X. Zhang}, {\it Phys. Rev.} {\bf D46}
(1992) {2190};
\item{E. Witten}, {\it Nucl. Phys.} {\bf B185} (1981) {513};
\item{J. C. Pati, M. Cveti\v{c}, H. Sharatchandra}, {\it Phys. Rev. Lett.}
{\bf 58} (1987) {851};
\item{K. S. Babu, J. C. Pati, H. Stremnitzer}, {\it Phys. Lett.}
{\bf B264} (1991) {347};
\item{E. Eichten, I. Hinchliffe, K. Lane, C. Quigg}, {\it Rev. Mod. Phys.}
{\bf 56} (1984) {579}.
\end{list}
\end{document}